\documentstyle[12pt,epsfig]{article}

\newcommand{\be}{\begin{equation}}
\newcommand{\ee}{\end{equation}}
\begin{document}  
\topmargin 0pt
\oddsidemargin=-0.4truecm
\evensidemargin=-0.4truecm
\renewcommand{\thefootnote}{\fnsymbol{footnote}}
\newpage
\setcounter{page}{0}
\begin{titlepage}   
\vspace*{-2.0cm}  
\begin{flushright}
FISIST/01-01/CFIF \\
hep-ph/0106201
\end{flushright}
\vspace*{0.5cm}
\begin{center}
{\Large \bf Global Analysis of Solar Neutrinos with Magnetic Moment and Solar Field Profiles} 
\vspace{1.0cm}

{\large 
Jo\~{a}o Pulido
\footnote{E-mail: pulido@beta.ist.utl.pt}}\\  
{\em Centro de F\'\i sica das Interac\c c\~oes Fundamentais (CFIF)} \\
{\em Departamento de Fisica, Instituto Superior T\'ecnico }\\
{\em Av. Rovisco Pais, P-1049-001 Lisboa, Portugal}\\
\end{center}
\vglue 0.8truecm

\begin{abstract}
A global statistical analysis of the pre-SNO solar neutrino data including the
rates and recoil electron spectrum in SuperKamiokande is made assuming 
the solar neutrino deficit to be resolved by the interaction of the neutrino 
magnetic moment with the solar magnetic field. Given the general characteristics
of the field profiles known to lead to the best event rate predictions, several 
specific choices of profiles are assumed and global fits investigated. 
Whereas previous studies revealed an excellent quality of the rate fits
within the magnetic moment solution to the solar neutrino problem, the
global fits are found to be only of reasonable quality and comparable to the 
global oscillation fits. This is related to the difficulty of any theoretical
model to provide at present an adjustable prediction to the spectral data.  
\end{abstract}
\end{titlepage}   


Whereas the apparent anticorrelation of the neutrino event rate with sunspot 
activity claimed long ago by the Homestake collaboration \cite{Homestake} 
remained unconfirmed by other experiments \cite{SuperK}, \cite{SAGE}, 
\cite{Gallex+GNO} and theoretical analyses \cite{Walther}, the magnetic moment 
solution to the solar neutrino problem is at present an important possibility 
to be explored in the quest for an explanation of the solar neutrino 
deficit. This is the idea, originally proposed by Cisneros \cite{Cisneros},
and later revived by Voloshin, Vysotsky and Okun \cite{VVO},  
that a large magnetic moment of the neutrino may interact with the magnetic 
field of the sun, converting weakly active to sterile neutrinos.
It now appears in fact that this deficit is energy dependent, in the sense
that neutrinos of different energies are suppressed differently. In order to 
provide an energy dependent deficit, the conversion mechanism from active to 
nonactive neutrinos must be resonant, with the location of the critical density 
being determined by the neutrino energy. Thus was developed the idea of the
resonant spin flavour precession (RSFP) proposed in 1988 \cite{LMA}. It involves
the simultaneous flip of both chirality and flavour consisting basically in the 
assumption that the neutrino conversion due to
magnetic moment and magnetic field takes place through a resonance inside matter 
in much the same way as matter oscillations \cite{MSW}. 
A sunspot activity related event rate 
in a particular experiment would hence imply that most of the neutrinos with energies 
relevant to that experiment have their resonances in the sunspot range. However,
the depth of sunspots is unknown (they may not extend deeper than a few hundred
kilometers) and the observed field intensity is too small in sunspots to allow
for a significant conversion. The anticorrelation argument has therefore lost 
its appeal for several years now.

Despite the absence of the anticorrelation argument,
several main reasons may be invoked to motivate RSFP and investigating its consequences
for solar neutrinos. In fact both RSFP and all oscillation scenarios indicate a 
drop in the survival probability from the low energy (pp) to the 
intermediate energy neutrino sector ($^7$Be, CNO, pep) and a subsequent moderate rise as 
the energy increases further into the $^8$B sector. The magnetic field profiles
providing good fits to the event rates from solar neutrino experiments typically show 
the characteristic of a sharp rise in intensity at some point in the solar 
interior, followed by a progressive moderate decrease \cite{PA}, \cite{Valle}. 
This is in opposite correspondence with the energy dependence of the probability  
in the sense that the strongest field intensities correspond to the smallest survival 
probabilities. Hence RSFP offers 
a unique explanation for the general shape of the probability, which naturally appears
as a consequence of the field profile. On the other hand, from solar physics and 
helioseismology such a sharp rise and peak field intensity is expected to occur along 
the tachoclyne, the region extending from the upper radiative zone to the lower 
convective zone, where the gradient of the angular velocity of solar matter is 
different from zero \cite{Parker}, \cite{ACT}. Furthermore, it has become 
clear \cite{GN}, \cite{PA}, \cite{Valle} that RSFP provides event rate fits 
from the solar neutrino experiments that are far better than all oscillation ones
\cite{BKS}, \cite{GG1}. Finally, there are recent claims in the literature for
evidence of a neutrino flux histogram \cite{SturrockS} containing two peaks, an
indication of variability pointing towards a nonzero magnetic moment of the neutrino.    

The aim of this paper is to present a global statistical analysis of three of the 
field profiles proposed in \cite{PA}, all obeying the general features described 
above, on the light of the updated standard solar model (BP 2000) \cite{BP2000} 
and the most recent data \cite{Homestake,SuperK,SAGE,Gallex+GNO}.  
This global analysis includes the rates and SuperKamiokande spectrum only, as no
day/night variations are expected for RSFP. 
Since the information on the SuperKamiokande total rate is already present 
in the flux of each spectral energy bin, this rate will not be included in the analysis,
thus following the same attitude as in ref. \cite {BKS1}. 
The SuperKamiokande energy spectrum used here corresponds to their new data with a range 
$E_e=(5.0-20)MeV$, 19 data points and 1258 days exposure time \cite{SuperK}. 
Moreover, all Gallium measurements (SAGE, Gallex and GNO) are combined in one single 
data point. For each profile the local minima of
the global $\chi^2$, defined as ${\chi^2}_{gl}={\chi^2}_{rates}+{\chi^2}_{spectrum}$,
are investigated in the ranges $\Delta m^2_{21}=(0-10^{-7})eV^2$ (mass squared difference
between neutrino flavours) and $B_0=(0-300)kG$ (value of the field at the peak). 
This $\Delta m^2$ range is related to neutrino resonances in the region from the upper radiation 
zone to the solar surface, that is, the whole region where a significant magnetic field 
is expected. The upper bound of 300 kG for the peak field ($B_0$) at the base of the 
convection zone is suggested by most authors \cite{Parker}, \cite{ACT}. The most poorly 
known solar neutrino flux, namely hep neutrinos, is taken as a free parameter, whereas 
the $^{8}B$ flux is fixed at its BP'2000 value \cite{BP2000}. The ${\chi^2}_{gl}$ local 
minima occur in the range $\Delta m^2_{21}=(0.7-2.0)\times10^{-8}eV^2$, 
with a peak field value from approximately 40kG up and $f_{hep}=
{\phi}(hep)_{(analysis)}/{\phi}(hep)_{(BP'98)} = 1.07-1.46$ where the BP'98 hep flux 
\cite{BP98} is taken as the reference one. The values of ${\chi^2}_{gl}$ at these local 
minima are found to be close to each other and around 20 for 18 d.o.f.. 
The neutrino magnetic moment is taken at $\mu_{\nu}=10^{-11}\mu_{B}$. 
This means that, since the order parameter is the product of $\mu_{\nu}$ 
by the magnetic field, for a value of the field at the peak near
300 kG, a suitable magnetic moment solution to the solar neutrino problem may exist
at the remarkably low value $\mu_{\nu}=1.4\times10^{-12}\mu_{B}$. The results are presented 
in terms of $2\sigma$ (95\%CL) and $3\sigma$ (99.73\%CL) contours on the $\Delta m^2_{21},B_0$ 
plane.

The profiles investigated are the following:

$Profile~(1)$
\be
B=0~~~,~~~x<x_R
\ee
\be
B=B_0\frac{x-x_R}{x_C-x_R}~,~x_{R}\leq x\leq x_{C}
\ee
\be
B=B_0\left[1-\frac{x-x_C}{1-x_C}\right]~ ,~x_{C}\leq x \leq 1
\ee
where $x$ is the fraction of the solar radius and $x_R=0.70$, $x_C=0.85$. 

$Profile~(2)$
\be
B=0~~~,~~~x<x_R
\ee
\be
B=B_0\frac{x-x_R}{x_C-x_R}~,~x_{R}\leq x\leq x_{C}
\ee
\be
B=B_0\left[1-\left(\frac{x-0.7}{0.3}\right)^2\right]~,~x_{C}<x\leq 1
\ee
with $x_R=0.65$, $x_C=0.75$.

$Profile~(3)$
\be
B=2.16\times10^3~~,~~x\leq 0.7105
\ee
\be
B=B_{1}\left[1-\left(\frac{x-0.75}{0.04}\right)^2\right]~,~0.7105<x<0.7483
\ee
\be
B=\frac{B_{0}}{\cosh30(x-0.7483)}~,~0.7483\leq x\leq 1
\ee
with $B_0=0.998B_1$.

The ratios of the RSFP to the SSM event rates $R^{th}_{Ga,Cl}$ are defined as before 
\cite{PA} and the recoil electron spectrum in SuperKamiokande is now
\be
R_{j}^{th}=\frac{\sum_{i}\int_{{E_e}_j}^{{E_e}_{j+1}}dE_e\int_{{E^{'}_e}_m}^{{E^{'}_e}_M}dE{'}\!\!_e
f(E{'}_e,E_e)\int_{E_m}^{E_M}dEf_i\phi_{i}(E)[P(E)\frac{d\sigma_{W}}{dT^{'}}+(1-P(E))
\frac{d\sigma_{\bar{W}}}{dT{'}}]}
{\sum_{i}\int_{{E_e}_j}^{{E_e}_{j+1}}dE_e\int_{{E^{'}_e}_m}^{{E^{'}_e}_M}dE{'}\!\!_e
f(E{'}_e,E_e)\int_{E_m}^{E_M}dE\phi_{i}(E)\frac{d\sigma_{W}}{dT^{'}}}
\ee
for 19 energy bins (j=1,...19) \cite{SuperK}.
Here $\phi_{i}(E)$ is the SSM neutrino flux for component i ($i=hep,~^8\rm{B}$) and the
factor $f_i$ is the ratio of the i-th flux relative to the SSM one. For $i=hep$ the
reference flux is taken to be the BP'98 one with $f_i$ as a free parameter, while for 
$i=^8\!\!B$, for which ${\phi_{^8\!B}}^{BP'98}={\phi_{^8\!B}}^{BP'2000}$, $f_i=1$ is assumed.
The quantity $f(E^{'}_e,E_e)$ is the energy resolution function \cite{Fukuda} of the detector in 
terms of the physical ($E^{'}_e$) and the measured ($E_e$) electron energy ($E_e=T+m_e$). 
The lower limit of $E^{'}_{e}$ is the detector threshold energy (${E^{'}_e}_{m}={E_e}_{th}$ 
with ${E_e}_{th}$= 5.0MeV) and the upper limit is evaluated from the maximum 
neutrino energy $E_M$ \cite{PA}
\be
T^{'}_{M}=\frac{2E^{2}_{M}}{m_e+2E_{M}}.
\ee 
For the lower \cite {PA} and upper \cite{Homepage} integration limits of the neutrino 
energy one has respectively
\be
E_m=\frac{T^{'}+\sqrt{T{^{'}}^{2}+2m_eT^{'}}}{2}~,~E_{M}=15MeV~(i=^{8}\!\!B)~,~E_{M}=18.8MeV~
(i=hep).
\ee
The weak differential cross sections appearing in equation (17) are given by 
\be
\frac{d\sigma_W}{dT}=\frac{{G_F}^2 m_e}{2\pi}[(g_V+g_A)^2+
(g_V-g_A)^2\left(1-\frac{T}{E}\right)^2-({g_V}^2-{g_A}^2)\frac{m_{e}T}{E^2}]
\ee
for $\nu_{e}e$ scattering, with $g_V=\frac{1}{2}+2sin^2\theta_{W}$, $g_A=\frac{1}{2}$. 
For $\bar\nu_{\mu} e$ and $\bar\nu_{\tau} e$ scattering,
\be
\frac{d\sigma_{\bar{W}}}{dT}=\frac{{G_F}^2 m_e}{2\pi}[(g_V-g_A)^2+
(g_V+g_A)^2\left(1-\frac{T}{E}\right)^2-({g_V}^2-{g_A}^2)\frac{m_e T}{E^2}]
\ee
with $g_V=-\frac{1}{2}+2sin^2\theta_{W}$, $g_A=-\frac{1}{2}$.

The partial event rates for each neutrino component in each experiment were 
taken from \cite{BP2000} and the solar neutrino spectra from Bahcall's homepage 
\cite{Homepage}. The contribution of the hep flux to both the Gallium and Homestake 
event rates was neglected. The $\chi^2$ analysis for the ratios of event rates and 
electron spectrum in SuperKamiokande was done following the standard procedure 
described in \cite{PA} \footnote{Here only the main definitions and differences are 
registered. For the calculational details we refer the reader to ref.\cite{PA}.}. The
validity of this procedure and alternative ones for solar neutrinos is discussed in 
refs. \cite {Garz+Giunti}, \cite{CSS}.     

The ratios of event rates to the SSM event rates and the recoil electron spectrum 
normalized to the SSM one, both denoted by $R^{th}$ in the following, were 
calculated in the parameter
ranges $\Delta m^2_{21}=(0-10^{-7})eV^2$, $B_0=(0-30)\times10^{4}G$ for
all magnetic field profiles and inserted in the $\chi^2$ definitions for the
rates and spectrum, 
\be
\chi^2_{rates}=\sum_{j_{1},j_{2}=1}^{2}({R}^{th}_{j_{1}}-{R_{j_{1}}}^{exp})\left[\sigma^{2}_{rates}
(tot)\right]^{-1}_{j_{1}j_{2}}({R}^{th}_{j_{2}}-{R_{j_{2}}}^{exp})
\ee
with (Ga=1, Cl=2)
\be
\chi^2_{sp}=\sum_{j_{1},j_{2}=1}^{19}({R}^{th}_{j_1}-R^{exp}_{j_1})[\sigma^{2}_{sp}(tot)]^{-1}
_{j_{1}j_{2}}({R}^{th}_{j_2}-R^{exp}_{j_2})\,.
\ee

The quantities $R^{exp}$ in eqs. (15), (16) are directly read from tables I, II respectively
and the total error matrices $\sigma^{2}(tot)$ are derived from the definitions given in
\cite{PA}, using \cite{Homepage} and the error bars in tables I, II. 
In performing the fitting one has 21 experiments (2 rates and 19 spectral data) and three
free parameters, namely the mass squared difference between neutrino flavours $\Delta m^2_{21}$,
the peak field value $B_0$ and the hep flux parametrized by $f_{hep}$, hence 18 degrees of 
freedom. With no correlations between the rates and spectral errors one has
for the global $\chi^2$, 

\be
{\chi^2}_{gl}={\chi^2}_{rates}+{\chi^2}_{spectrum}.
\ee

For each of the three profiles, eqs.(1)-(9), the local minima of $\chi^2_{gl}$ in the above
parameter ranges were first investigated. Five such minima were found for the equilateral
triangle, profile 1, all with $\chi^2_{gl}/18d.o.f.\simeq 20$ (see table III). For the second 
profile, seven local minima occur with a $\chi^2_{gl}/18d.o.f.$ ranging from 20 to 24 
(see table IV) and for the third 
one only two appear with $\chi^2_{gl}/18d.o.f.=21-22$ (see table V). Each of these 
minima is obtained for an hep flux from 1.07 to 1.46 of its BP'98 value as also shown.
In view of the closeness of the $\chi^2$ values for all these fits with the possible exception 
of three of them, it is hard to tell which values one should expect the data to prefer.
It is clear on the other hand that, since the value of the field at the peak may be as
large as 300 kG and this analysis was done using $\mu_{\nu}=10^{-11}\mu_B$, a value 
$B_0=O(40kG)$ (table III) means the existence of a possible solution to the solar neutrino problem with 
$\mu_{\nu}$ as low as $1.4\times10^{-12}\mu_B$ in consistency with astrophysical bounds 
\cite{magmo}. Moreover, the fits with the larger $B_0$ 
values are disfavoured, as they would imply a value of the neutrino magnetic moment close to 
$10^{-11}\mu_B$, in conflict with most astrophysical bounds \cite{magmo}.

The allowed physical regions and stability of these fits were also analysed. The results
are presented in terms of $2\sigma$ and $3\sigma$
contours (95\% and 99.73\% CL respectively) in the $\Delta m^2_{21}$, $B_0$ plane. These 
contours enclose the regions for which $\chi^2 \leq \chi^2_{min}+\Delta \chi^2$ with, for
18 d.o.f., $\Delta \chi^2=28.87$ (95\%CL) and $\Delta \chi^2=39.17$ (99.73\%CL) \cite{RPP}. 
They are displayed in figs.1, 2, 3 for each of the profiles along with the local $\chi^2$
minima. The dashed lines denote $2\sigma$ and the full lines denote $3\sigma$ contours.  
The local minima chosen for reference, that is, the ones for which the contours are defined, 
are numbers II (profile 1), VI (profile 2) and XIII (profile 3). In view of the closeness
of the $\chi^2$ values at the minima, their choice is irrelevant in terms of the difference
among contours belonging to different minima: the discrepancies affecting  
the parameters $\Delta m^2_{21}$, $B_0$ are hardly perceptible on the scale
of the plots. Finally, the contours were worked out for an hep flux ($f_{hep}$) fixed to
its value at the chosen minimum of $\chi^2$. 

As an example, the predicted spectrum (eq.(10)) for profile 3 
is shown in fig. 4 at the global best fit (XIII) superimposed on the SuperKamiokande data. 
For this case $\chi^2_{sp}=21.11$. The moderate rise occuring in the theoretical
curve for $E_e \geq 12 MeV$ is the effect of the hep neutrinos. It is clear from this 
figure the inherent difficulty in fitting any model prediction to the experimental data, 
a fact obviously also present in statistical analyses of oscillations.  

To conclude, the analysis of prospects for the magnetic moment solution to the solar
neutrino problem reveals global fits of reasonable quality. Previous analyses \cite{PA},
\cite{Valle},\cite{GN} unambiguously indicated from the part of the data on rates a preference for profiles
with a steep rise across the bottom of the convection zone in which vicinity they
reach a maximum, followed by a more moderate decrease up to the solar surface.
This leads to an excellence of rate fits alone which is not shared by the oscillation 
rate fits. Interestingly enough, this class of profiles is the most consistent one
with solar physics and helioseismology \cite{Parker,ACT}. On the other hand the global 
analysis presented here for rates and spectral data shows a quality of fits comparable 
to the oscillation ones. This originates from the relative poor quality of the spectrum 
fits caused by the shape of a function for which no theoretical model is able 
to provide at present a properly adjustable prediction. 

In the present analysis only time averaged data were considered and a fitting was made 
to a time constant profile 'buried' in the solar interior. If, on the contrary, the active
neutrino flux turns out to be time dependent, a situation most likely to be interpreted 
through the magnetic moment solution with a time dependent interior field, the present 
approach is obviously inadequate. Averaging the event rates over time implies disregarding 
possible information in the data which otherwise is available if different periods of time 
are considered \cite{SturrockS}. The robustness of such a procedure will greatly improve 
with the accumulation of more data.

\newpage


\begin{center}
\begin{tabular}{lcccc} \\ \hline \hline
Experiment &  Data      &   Theory   &   Data/Theory  &  Reference \\ \hline
Homestake  &  $2.56\pm0.16\pm0.15$ & $7.7\pm^{1.3}_{1.1}$ & $0.332\pm0.05$ & 
\cite{Homestake} \\
Ga     &  $74.7\pm5.13$ & $129\pm ^8_6$ & $0.59\pm0.06$ &
\cite{Gallex+GNO},\cite{SAGE} \\
SuperKamiokande&$2.4\pm0.085$ &
$5.15\pm^{1.0}_{0.7}$&$0.465\pm0.052$& \cite{SuperK}\\ \hline
\end{tabular}
\end{center}

{Table I - Data from the solar neutrino experiments. Units are SNU for
Homestake and Gallium and $10^{6}cm^{-2}s^{-1}$ for SuperKamiokande. The
result for Gallium is the combined one from SAGE and Gallex+GNO.}

\begin{center}
\begin{tabular}{cc} \\ \hline \hline
Energy bin (MeV)& $R_j^{exp}$ \\ \hline
$5<E_e<5.5$     & $0.436\pm0.046$ \\
$5.5<E_e<6$     & $0.438\pm0.024$ \\
$6<E_e<6.5$     & $0.435\pm0.019$  \\
$6.5<E_e<7$     & $0.438\pm0.015$ \\ 
$7<E_e<7.5$     & $0.463\pm0.015$ \\ 
$7.5<E_e<8$     & $0.483\pm0.016$ \\
$8<E_e<8.5$     & $0.465\pm0.017$ \\
$8.5<E_e<9$     & $0.438\pm0.017$ \\
$9<E_e<9.5$     & $0.450\pm0.018$ \\
$9.5<E_e<10$    & $0.455\pm0.019$ \\
$10<E_e<10.5$   & $0.442\pm0.021$ \\
$10.5<E_e<11$   & $0.407\pm0.022$ \\
$11<E_e<11.5$   & $0.455\pm0.026$ \\
$11.5<E_e<12$   & $0.423\pm0.028$ \\
$12<E_e<12.5$   & $0.422\pm0.033$ \\
$12.5<E_e<13$   & $0.481\pm0.041$ \\
$13<E_e<13.5$   & $0.431\pm0.047$ \\
$13.5<E_e<14$   & $0.603\pm0.065$ \\
$14<E_e<20$     & $0.493\pm0.049$ \\ \hline
\end{tabular}
\end{center}

{Table II - Spectral energy bins in SuperKamiokande (1258 days) and the corresponding
ratio of the experimental to the SSM event rate \cite{SuperK}.}

\newpage

\begin{center}
\begin{tabular}{ccccc}\\ \hline \hline
Fit & $\Delta m^2_{21}(eV^2)$ & $B_0~(G)$ & $f_{hep}$ & $\chi^2_{gl}$/18d.o.f.\\ \hline
I        & $6.68\times10^{-9}$ & $4.1\times10^{4}$ & $1.21$ & $21.32$\\
II        & $6.75\times10^{-9}$ & $6.73\times10^{4}$  &$1.19$  & $21.49$\\
III        & $8.26\times10^{-9}$ & $1.34\times10^{5}$  &$1.36$  & $20.22$\\ 
IV        & $7.5\times10^{-9}$ & $1.7\times10^{5}$  & $1.19$  &$21.37$ \\
V        & $1.0\times10^{-8}$ & $2.32\times10^{5}$ & $1.46$ &$19.95$  \\ \hline
\end{tabular}
\end{center}  

{Table III - Local minima of $\chi^2_{gl}$ for profile 1 (eqs.(1)-(3)). The reference for
which the $2\sigma$ and $3\sigma$ contours are determined and drawn in fig.1 is fit II.}
 
\begin{center}
\begin{tabular}{ccccc}\\ \hline \hline
Fit & $\Delta m^2_{21}(eV^2)$ & $B_0~(G)$ & $f_{hep}$ & $\chi^2_{gl}$/18d.o.f.\\ \hline
VI        & $1.22\times10^{-8}$ & $9.4\times10^{4}$ & $1.24$ & $21.08$\\
VII        & $1.21\times10^{-8}$ & $1.22\times10^{5}$  &$1.07$  & $24.29$\\
VIII        & $1.28\times10^{-8}$ & $1.7\times10^{5}$  &$1.31$  & $20.33$\\ 
IX        & $1.3\times10^{-8}$ & $1.99\times10^{5}$  & $1.09$  &$23.50$ \\
X        & $1.38\times10^{-8}$ & $2.46\times10^{5}$  & $1.43$  &$19.92$ \\
XI        & $1.41\times10^{-8}$ & $2.76\times10^{5}$ & $1.11$ &$23.25$  \\ 
XII        & $1.56\times10^{-8}$ & $3.23\times10^{5}$ & $1.45$ &$19.88$  \\ \hline
\end{tabular}
\end{center}  

{Table IV - Local minima of $\chi^2_{gl}$ for profile 2 (eqs.(4)-(6)). The reference for
which the $2\sigma$ and $3\sigma$ contours are determined and drawn in fig.2 is fit VI.}

\begin{center}
\begin{tabular}{ccccc}\\ \hline \hline
Fit & $\Delta m^2_{21}(eV^2)$ & $B_0~(G)$ & $f_{hep}$ & $\chi^2_{gl}$/18d.o.f.\\ \hline
XIII        & $1.36\times10^{-8}$ & $1.04\times10^{5}$ & $1.26$ & $21.24$\\
XIV        & $1.82\times10^{-8}$ & $1.90\times10^{5}$  &$1.17$  & $22.71$\\ \hline
\end{tabular}
\end{center}  

{Table V - Local minima of $\chi^2_{gl}$ for profile 3 (eqs.(7)-(9)). The reference for
which the $2\sigma$ and $3\sigma$ contours are determined and drawn in fig.3 is fit XIII.}


\newpage

\centerline{\large Figure captions}

\noindent
Fig. 1.
Fits corresponding to the local minima of $\chi^2_{gl}$ for profile 1 (stars) and the
95\% and 99.73\%CL contours (dashed and solid lines respectively) with respect to fit
II. Since the values of $\chi^2_{gl}$ at these minima lie close to each other (see 
table III), the contours may be regarded as actually enclosing the common $2\sigma$ and 
$3\sigma$ allowed areas on the scale of the plot for all fits. 

\noindent
Fig. 2.
Same as fig.1 for profile 2. The reference fit is VI. 

\noindent
Fig. 3.
Same as fig.1 for profile 3. The reference fit is XIII.

\noindent
Fig. 4.
The theoretical prediction of the SuperKamiokande recoil electron spectrum (solid line)
superimposed on the data set \cite{SuperK} (1258 days) for profile 3 at the global
fit XIII. Here the value of $\chi^2$ for the spectrum is $\chi^2_{sp}=21.11$.
\newpage

\begin{figure}
\mbox{\psfig{figure=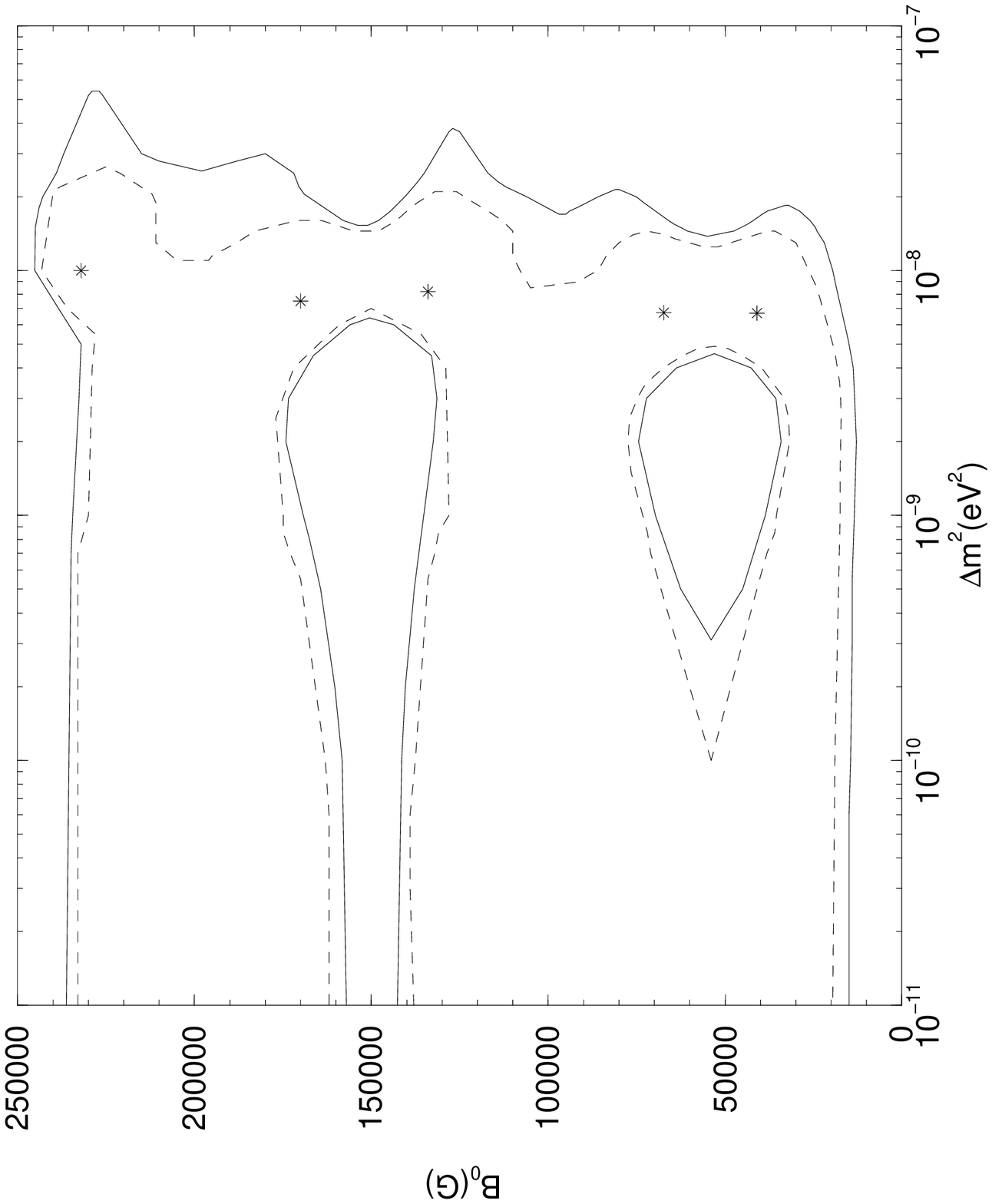,width=19cm}}
\centerline{\mbox{Fig. 1.}}
\end{figure}

\newpage

\begin{figure}
\mbox{\psfig{figure=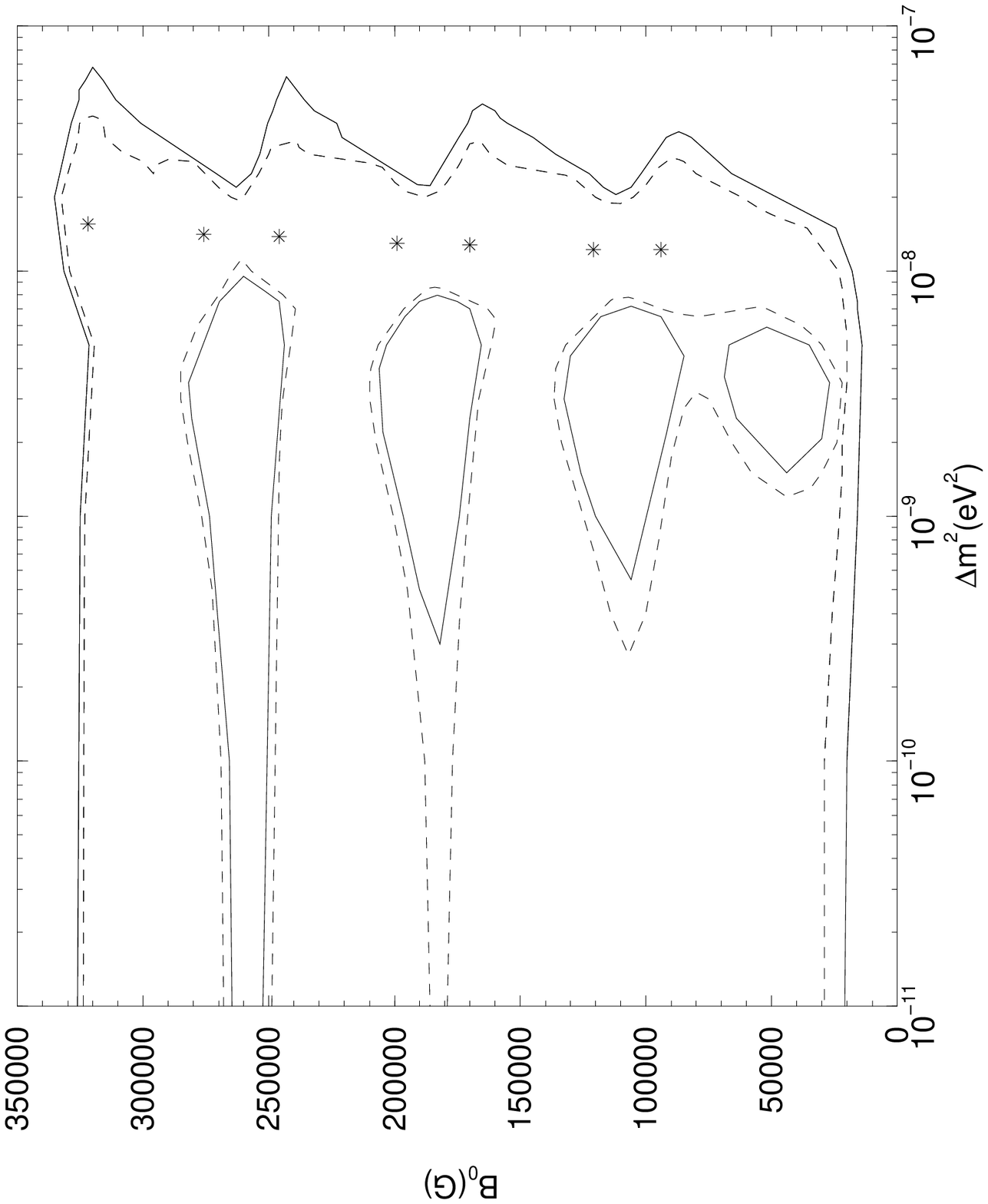,width=16cm}}
\centerline{\mbox{Fig. 2.}}
\end{figure}

\newpage

\begin{figure}
\mbox{\psfig{figure=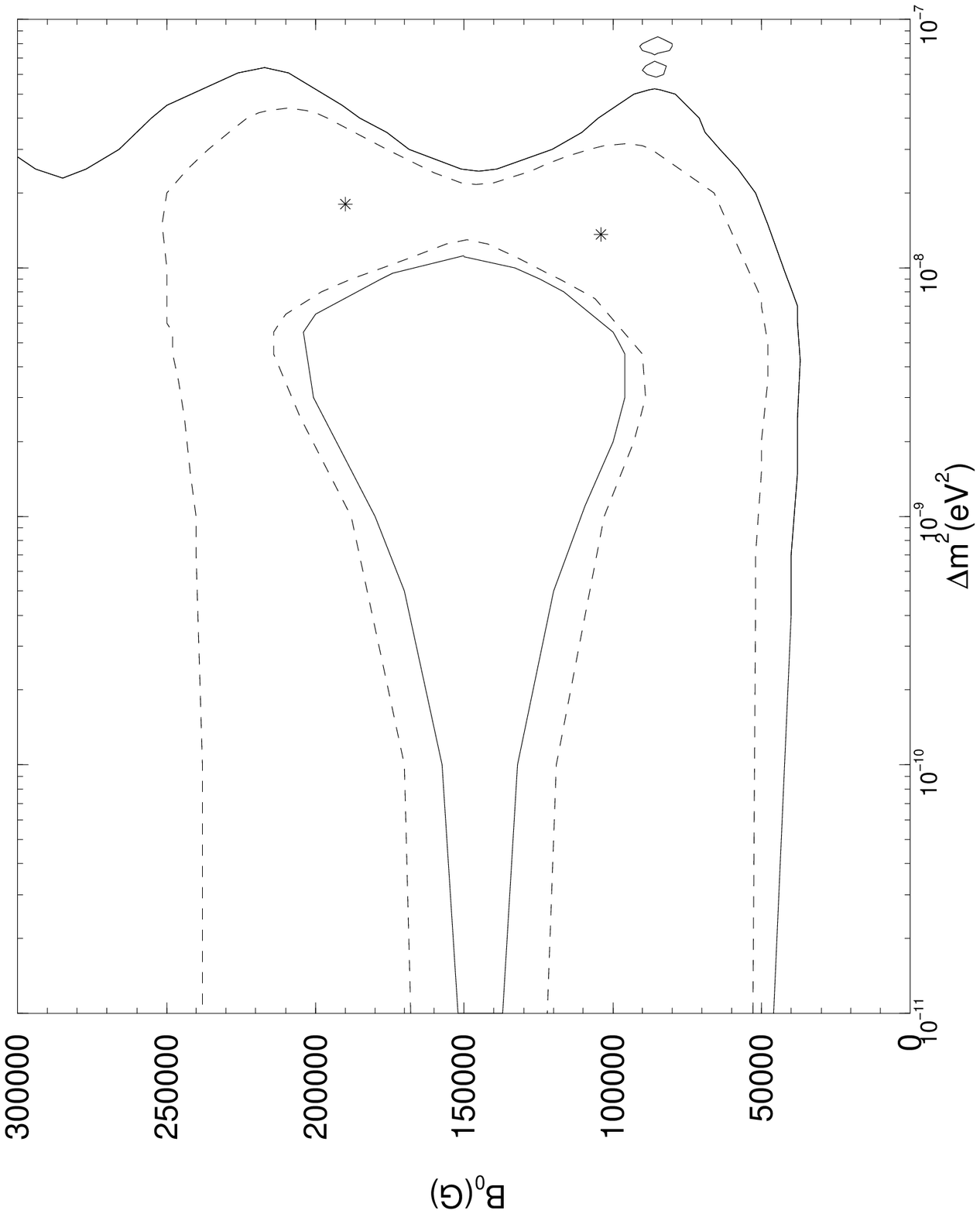,width=16cm}}
\centerline{\mbox{Fig. 3.}}
\end{figure}

\newpage

\begin{figure}
\mbox{\psfig{figure=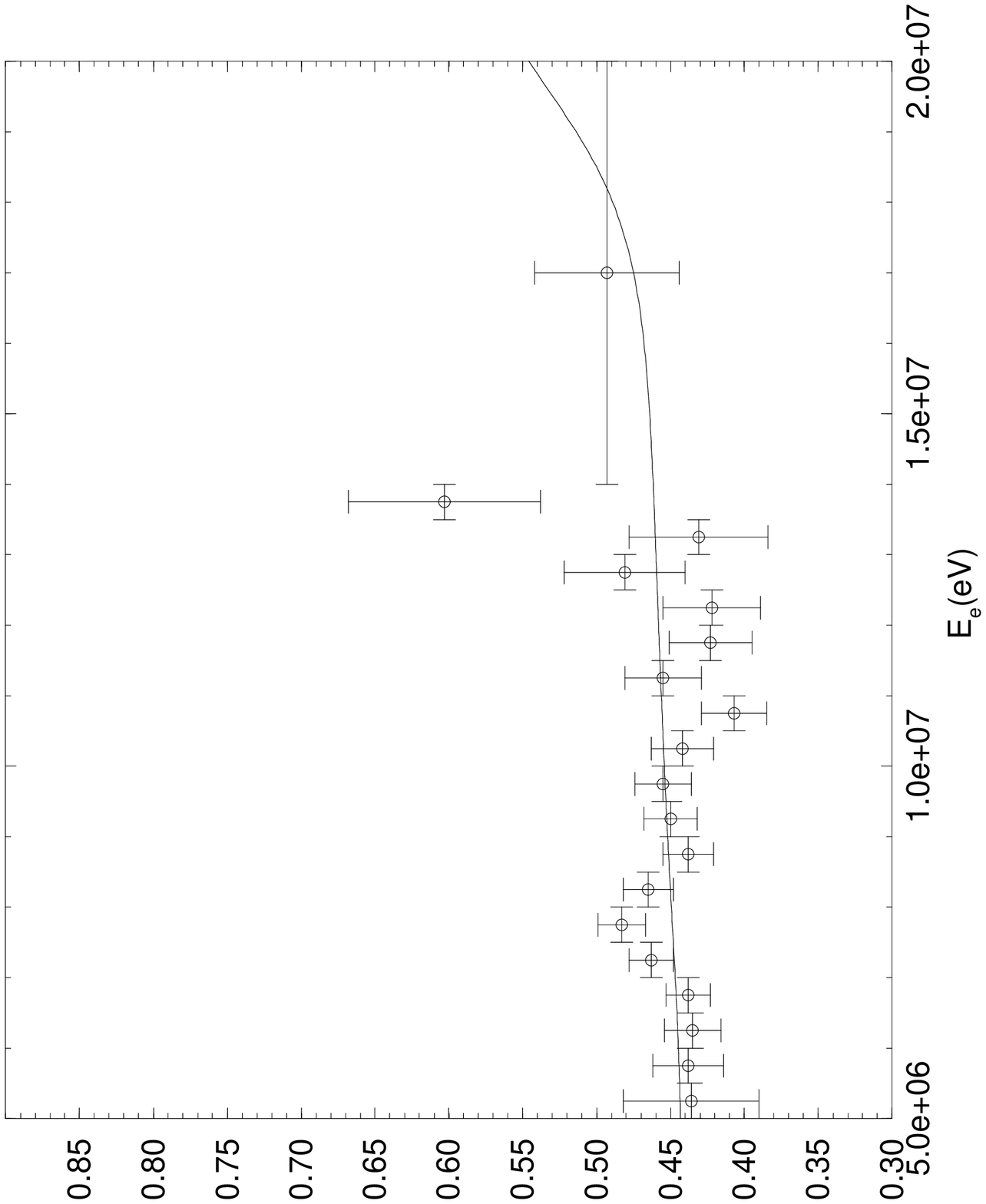,width=17cm}}
\centerline{\mbox{Fig. 4.}}
\end{figure}

\end{document}